\documentclass[conference]{IEEEtran}
\IEEEoverridecommandlockouts
\usepackage{cite}
\usepackage{amsmath,amssymb,amsfonts}
\usepackage{algorithmic}
\usepackage{graphicx}
\usepackage{textcomp}
\usepackage{xcolor}
\usepackage{caption}
\usepackage{lipsum}
\usepackage{subcaption}
\usepackage{upgreek}
\usepackage{booktabs} 
\def\BibTeX{{\rm B\kern-.05em{\sc i\kern-.025em b}\kern-.08em
    T\kern-.1667em\lower.7ex\hbox{E}\kern-.125emX}}
\begin{document}

\title{Measurement-Based Validation of Geometry-Driven RIS Beam Steering in Industrial Environments\\

\thanks{This work was supported in part by the German Federal Ministry of
Research, Technology and Space (BMFTR) in the course of the 6GEM+ Transfer Hub
under grant 16KIS2411 and in part by the German Research Foundation (DFG) under Project–ID 287022738 TRR 196 for Project S03.}
}

\author{
    \IEEEauthorblockN{Adam Umra$^\star$, Simon Tewes$^\star$, Niklas Beckmann$^\dagger$, Niels König$^\dagger$, Aydin Sezgin$^\star$, Robert Schmitt$^\dagger$}
    \IEEEauthorblockA{$^\star$Ruhr University Bochum, Germany\\
    {\{adam.umra, simon.tewes, aydin.sezgin\}@rub.de}\\
    $^\dagger$Fraunhofer Institute for Production Technology, Germany\\
    {\{niklas.beckmann, niels.koenig, robert.schmitt\}@ipt.fraunhofer.de}}
}

\maketitle

\begin{abstract}
Reconfigurable intelligent surfaces (RISs) offer programmable control of radio propagation for future wireless systems. For configuration, geometry-driven analytical approaches are appealing for their simplicity and real-time operation, but their performance in challenging environments such as industrial halls with dense multipath and metallic scattering is not well established. To this end, we present a measurement-based evaluation of geometry-driven RIS beam steering in a large industrial hall using a 5~GHz RIS prototype. A novel RIS configuration is proposed in which four patch antennas are mounted in close proximity in front of the RIS to steer the incident field and enable controlled reflection. For this setup, analytically computed, quantized configurations are implemented. Two-dimensional received power maps from two measurement areas reveal consistent, spatially selective focusing. Configurations optimized near the receiver produce clear power maxima, while steering to offset locations triggers a rapid 20--30~dB reduction. With increasing RIS--receiver distance, elevation selectivity broadens due to finite-aperture and geometric constraints, while azimuth steering remains robust. These results confirm the practical viability of geometry-driven RIS beam steering in industrial environments and support its use for spatial field control and localization under non-ideal propagation.
\end{abstract}


\section{Introduction}

Industrial wireless networks are facing increasingly stringent performance requirements, driven by applications that require ultra-high reliability and very low end-to-end delay with tight latency bounds~\cite{luvisotto2019Indust}. In large production halls, radio propagation is strongly shaped by dense multipath, extensive metallic infrastructure, and highly reflective surfaces. Careful radio planning and site-specific optimization can mitigate many effects in static environments, but reconfigurable and matrix manufacturing~\cite{mohr2024Matrix} create dynamic layouts that challenge conventional planning. In such environments, link characteristics may vary frequently over time, reducing the predictability and long-term effectiveness of static beamforming and pre-planned wireless configurations~\cite{noor2023Smart}. Accordingly, there is growing interest in methods that provide dynamic, environment-level control over wireless propagation.

\emph{Reconfigurable intelligent surfaces} (RISs) have gained traction as a means to shape electromagnetic propagation in future sixth-generation (6G) wireless systems~\cite{basar2019RISaccess,renzo2020RIS}. By enabling programmable control over the reflection properties of large surfaces, RISs can improve coverage, robustness, and energy efficiency~\cite{wu2020RIS,weinberger2022RISRAN}, particularly in environments that are difficult to influence through conventional infrastructure. Interest in RIS-assisted systems also extends beyond communication to sensing-related applications such as localization, spatial field shaping, and environment-aware signal processing~\cite{magbool2025RIS,weinberger2024UWBris, umra2025hardwareefficientcognitiveradarmultitarget}.

A central practical requirement is the ability to compute RIS configurations efficiently. In particular, geometry-based analytical channel models are attractive in this context because they provide low-complexity and real-time-capable optimization with clear physical interpretability. However, such models are typically derived under idealizing assumptions (e.g., free-space propagation, simplified reflection behavior, incomplete environment knowledge). Industrial shop floors are often dominated by non-line-of-sight (NLOS) components, uncontrolled scattering, and dense multipath propagation. In such environments, it remains unclear whether configurations derived purely from geometric models can deliver consistent performance in practice. In particular, it is uncertain whether these configurations still provide spatially localized and reliable control of the received field once deployed on real hardware.

This paper investigates this question through a measurement-based study in a representative industrial production hall. We focus on the end-to-end effectiveness of model-driven, geometry-based RIS beam steering under strongly non-ideal radio propagation. Specifically, we evaluate whether analytically computed RIS configurations yield spatially selective and repeatable field focusing when deployed directly on a hardware RIS prototype in the presence of severe multipath and metallic scattering.

\subsection{Related Work}

Several experimental studies have examined the validity of analytical RIS channel models using hardware prototypes in controlled indoor environments. In~\cite{weinberger2023geo}, a geometry-based RIS channel model is evaluated in an indoor setting by comparing analytically optimized configurations with measurements, revealing deviations attributable to hardware impairments and modeling simplifications. Similarly,~\cite{weinberger2024ValRIS} investigates specific model assumptions through measurements in an anechoic chamber, showing that non-perpendicular reflection angles introduce attenuation and phase effects that are not captured by ideal free-space models.

In contrast to these works, which primarily validate or refine analytical models under controlled indoor propagation conditions, the present work assesses the practical effectiveness of geometry-based RIS beam steering in a realistic, large industrial environment. The emphasis is on experimentally quantifying spatial controllability and repeatability using measured two-dimensional field distributions.

\subsection{Contributions}

This paper presents a measurement-based assessment of model-driven RIS beam steering in an industrial production hall. We conduct an extensive measurement campaign using a 5~GHz RIS prototype with binary phase control and a densely discretized spatial measurement grid. We introduce a new transmitter (Tx)–RIS hardware configuration in which four patch antennas are mounted directly in front of the RIS to illuminate the surface and steer the transmitted field toward it, enabling controlled reflection and wavefront manipulation for the experiments. For each grid position, RIS configurations are computed from geometry using an analytical method and deployed without feedback or environment-specific calibration. Based on the resulting two-dimensional received power maps, we characterize the degree of spatial field control achievable with RIS beam steering under realistic industrial radio conditions. The results show that geometry-based beam steering can produce effective and repeatable spatial focusing despite severe multipath and metallic scattering. While increased RIS-to-receiver distance introduces geometry-induced limitations in elevation resolution, robust azimuthal steering performance is maintained. These findings support the practical viability of model-driven RIS optimization in industrial settings and motivate its use for localization in future industrial wireless systems.

\section{System and Channel Model}\label{sec:sys}

We consider a system composed of a single-antenna Tx, a RIS, and a single-antenna receiver (Rx). The RIS consists of $N$ modules, each equipped with $M$ passive reflecting elements (see Fig.~\ref{fig:ris_close}), resulting in a total of $NM$ RIS elements. The Tx illuminates the RIS, and the signals reflected by the RIS elements are subsequently received at the Rx.

By properly configuring the phase responses of the RIS elements, the reflected signal components can be coherently combined at the receiver. Under this assumption, the resulting effective baseband channel between the Tx, RIS and Rx is given by
\begin{align}
    h^\mathsf{eff} = \sqrt{G_T G_R} \sum_{m=1}^{NM} h_m \theta_m g_m ,
    \label{eq:heff_4tx}
\end{align}
where $h_m \in \mathbb{C}$ denotes the channel coefficient between the transmit antenna and the $m$-th RIS element, and $g_m \in \mathbb{C}$ represents the channel coefficient between the $m$-th RIS element and the receiver. The terms $G_T$ and $G_R$ denote the antenna gains at the transmitter and receiver, respectively. Each RIS element applies a complex-valued reflection coefficient $\theta_m = A_m(\varphi_m)e^{j\varphi_m}$, where $\varphi_m$ is the adjustable phase shift and $A_m(\varphi_m) \in [0,1]$ models the phase-dependent reflection amplitude.

For analytical convenience, we define the cascaded channel corresponding to the $m$-th RIS element as
\begin{align}
    h^\mathsf{casc}_m = h_m g_m ,
    \label{eq:casc_chan}
\end{align}
which captures the combined propagation effects of the Tx--RIS and RIS--Rx links. To explicitly characterize these cascaded channel components, we exploit the geometry of the considered system. Let $d^h_m$ denote the distance between the transmit antenna and the $m$-th RIS element, and let $d^g_m$ denote the distance between the $m$-th RIS element and the receiver. According to the free-space propagation model in \cite{goldsmith2005WC}, the individual channel coefficients can be expressed as
\begin{align}
    h_m = \frac{c}{4\pi f d^h_m} e^{j\frac{2\pi}{\lambda} d^h_m},
\end{align}
and
\begin{align}
    g_m = \frac{c}{4\pi f d^g_m} e^{j\frac{2\pi}{\lambda} d^g_m},
\end{align}
where $c$ denotes the speed of light, $f$ is the carrier frequency, and $\lambda$ is the corresponding wavelength. Consequently, the cascaded channel coefficient associated with the $m$-th RIS element is given by
\begin{align}
    h^\mathsf{casc}_m =
    \left( \frac{c}{4\pi f d^h_m} e^{j\frac{2\pi}{\lambda} d^h_m} \right)
    \left( \frac{c}{4\pi f d^g_m} e^{j\frac{2\pi}{\lambda} d^g_m} \right).
    \label{eq:chanModel_4tx}
\end{align}
\vspace{1pt}

\section{Experimental Validation}\label{sec:exp}

\subsection{Setup}

\begin{figure}[htbp]
    \centering
    \subfloat[Close-up view of the Tx \& RIS.]{%
        \includegraphics[width=0.75\linewidth]{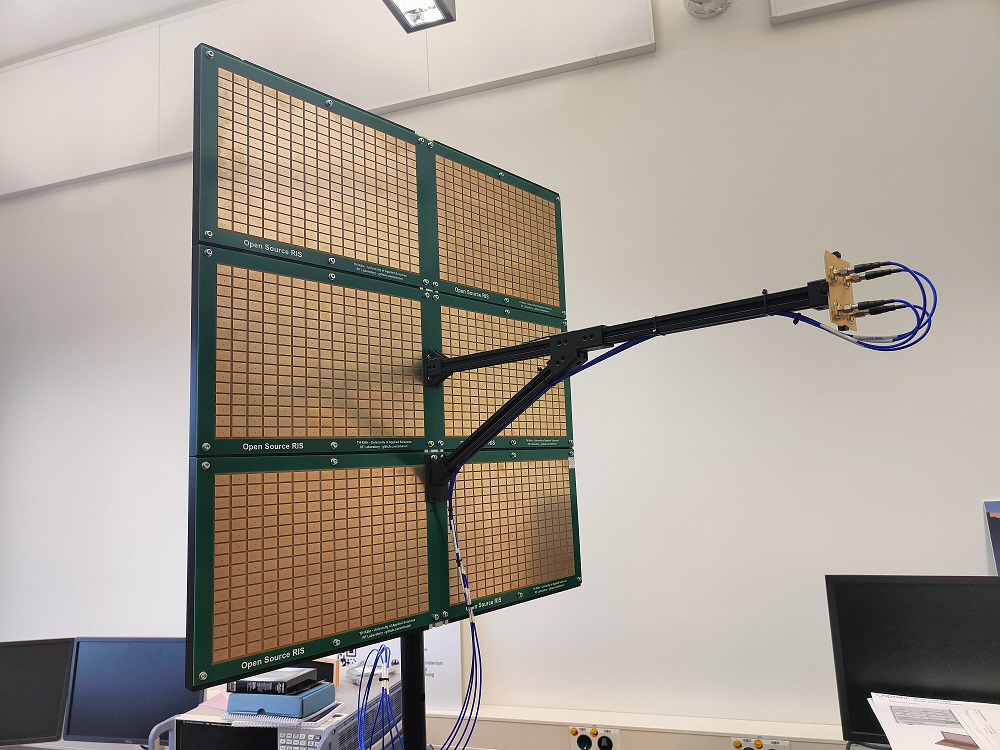}
        \label{fig:ris_close}
    }\\
    \subfloat[Tx and RIS mounted on a tripod with the Multistation positioned in front of the setup.]{%
        \includegraphics[width=0.75\linewidth]{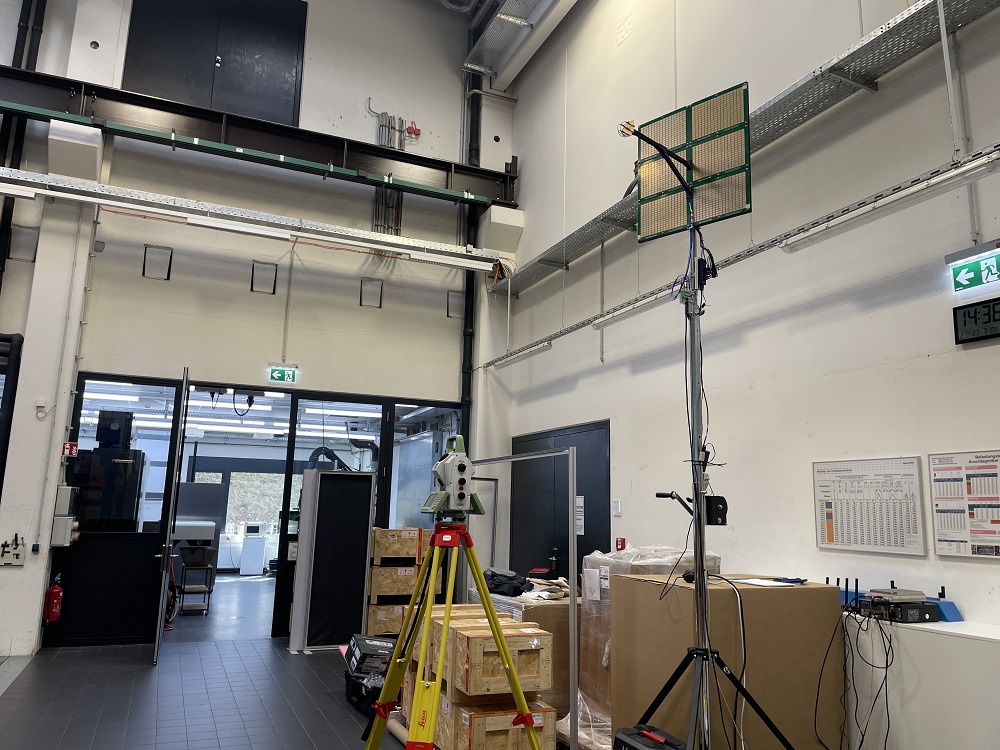}
        \label{fig:ris_far}
    }
    \caption{Overview of the experimental setup at the Fraunhofer IPT production hall, including the TX, RIS, and the Leica Multistation.}
    \label{fig:ris}
\end{figure}

\begin{figure*}[htbp]
    \centering
    \includegraphics[width=1\linewidth]{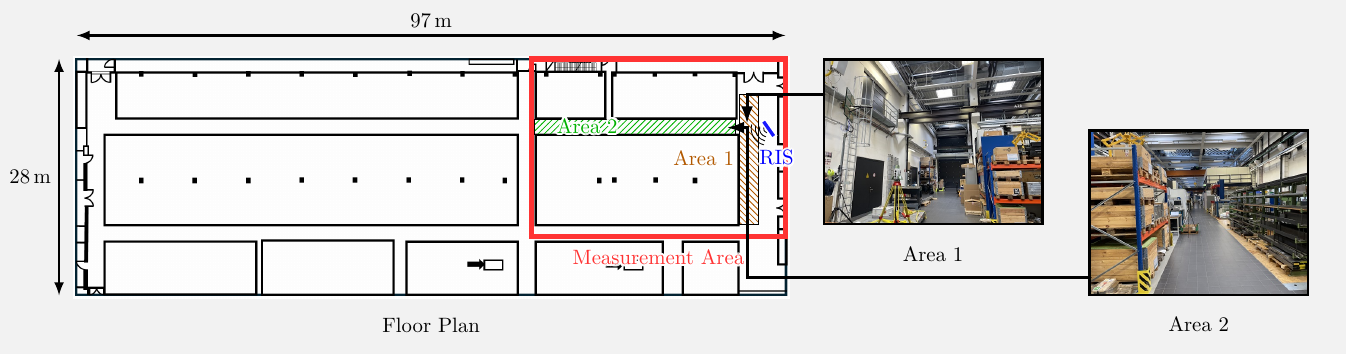}
    \caption{Floor plan of the production hall at Fraunhofer IPT in Aachen. The area where measurements were conducted is highlighted by a red square. The considered areas are marked with green and brown shading, and the RIS position is marked with a blue rectangle.}
    \label{fig:floor}
\end{figure*}

An experimental scenario was implemented using a RIS prototype operating in the $5\,\mathrm{GHz}$ band. A close-up view of the employed RIS is shown in Fig.~\ref{fig:ris_close}. Each RIS module has physical dimensions of $360\,\mathrm{mm} \times 247\,\mathrm{mm}$ and is described in detail in~\cite{heinrichs2023RIS}. In the considered setup, $N=6$ RIS modules were mounted on a tripod and arranged in a $3 \times 2$ planar configuration (see Fig.~\ref{fig:ris_far}). The measurements were conducted in a real-world large-scale production environment at Fraunhofer IPT~\cite{FraunhoferIPT_5GIndustryCampusEurope} in Aachen, Germany. The corresponding floor plan is depicted in Fig.~\ref{fig:floor}. The factory hall has a length of $97.0\,\mathrm{m}$, a width of $28.5\,\mathrm{m}$, and an approximate height of $10\,\mathrm{m}$, resulting in a total shop-floor area exceeding $2{,}700\,\mathrm{m}^2$. The measurements focused on the upper-right region of the hall, as highlighted with a red square in Fig.~\ref{fig:floor}.
Each RIS module consists of $M=256$ unit cells arranged in a $16 \times 16$ grid. Every unit cell is equipped with an RF switch that enables phase control. The RIS operates as a binary phase-switching surface, providing a $180^\circ$ phase shift at the designed carrier frequency ($5.15-5.875 \,\mathrm{GHz}$).
As illustrated in Fig.~\ref{fig:ris_close}, four patch antennas are strategically integrated directly in front of the RIS, forming a compact illumination architecture. Rather than relying on a distant feeder, these antennas directly illuminate the surface and independently steer their transmitted signals toward the RIS, enabling controlled reflection and wavefront manipulation. Signal transmission was realized using a USRP B200 software-defined radio (SDR) from Ettus Research as the transmitter, while signal reception was carried out with a Signal Hound SM435 spectrum analyzer. On the receiver side, a standard gain horn antenna was employed to ensure directional reception and sufficient antenna gain during the measurements. 

A Leica multistation (surveying instrument for precise position measurement) was used to survey two distinct areas in front of the RIS as marked with brown (Area 1) and green (Area 2) shading in Fig.~\ref{fig:floor}. The measurement areas were discretized into a set of grid points. To demonstrate the ability of the RIS to control the electromagnetic field distribution, the RIS configuration was optimized individually for each grid point. A model-based optimization approach based on the previously introduced system model was employed to maximize the received power at the selected grid point. The details of the optimization procedure are described in the following subsection. This optimization process was performed for all grid points in the measurement area.

\subsection{Model-based Optimization}

To enable real-time optimization of the RIS configurations at each grid point, we employ the channel model described in Section~\ref{sec:sys}. As noted previously, the experimental setup uses four patch antennas. For compatibility with the system model, we approximate this antenna array as a single equivalent antenna, whose location is defined as the geometric center of the four patch antennas. Based on the grid point position, we can generate the current geometry of the system to determine the channels between the grid point and the reflecting elements of the RIS. Thus, given the channel values, we are able to compute a beam to this grid point and round the values to the closest possible switch state. Given the absence of a direct link between Tx and Rx, we account for various phase values in the effective channel to optimize the RIS state for achieving the most efficient configuration.

In order to guarantee real-time capability, we compute the RIS configuration analytically, expressed as 
\begin{align}
    \varphi_m^*(\vartheta_t) = \vartheta_t - \varphi'_m,
\end{align}
where $\vartheta_t$ denotes an arbitrary value for the desired phase at the receiver, due to the absence of a direct Tx-Rx link. Accordingly, the cascaded channel phase $\varphi'_m$ is given by
\begin{align}
\varphi'_m
= \frac{2\pi}{\lambda} d^h_m
+ \frac{2\pi}{\lambda} d^g_m.
\label{eq:casc_phase}
\end{align}

Although the flexibility in selecting $\vartheta_t$ allows us to iterate through any number of evenly-distributed phase values $\vartheta_t$, we reduce the maximum number of allowed values to $T = 4$. This achieves a balanced tradeoff between the quality of the RIS-enabled link and the time needed in order to determine the optimal configuration. After determining the analytical solution for the $T$ phase values, we round the continuous phase shifts at each reflecting element to the nearest possible binary switching state of the RIS prototype. For these prototype-deployable RIS configurations, we choose the best performing one by assessing and comparing the simulated performances of the resulting RIS-facilitated links. This process can be represented mathematically as
\begin{align}
h_t^\mathsf{eff} = \hspace{-0.1cm} \sum_{m=1}^{N M} {h}^\mathsf{casc}_m A_m(\text{rd}(\varphi_m^*(\vartheta_t))) e^{j \text{rd}(\varphi_m^*(\vartheta_t))},\,  \forall \vartheta_t \in \mathsf{\Theta},\label{optEq}
\end{align}

\begin{align}
 \text{with} \quad A_m(\tau)&= \begin{cases} 0.5012 (-3\text{dB}) , &\text{if } \tau = \pi \\ 1 , &\text{otherwise}\end{cases}&,\\
 \text{rd}(\tau) &= \begin{cases} \pi , &\text{if } \frac{\pi}{2} \leq\tau < \frac{3\pi}{2} \\ 0, &\text{otherwise}\end{cases}&, \label{rdFun}\\[-1pt]
 \mathsf{\Theta} &= \left\{ 0, \frac{\pi}{2}, \pi, \frac{3\pi}{2} \right\}, |\mathsf{\Theta}|=T &,\label{Theta} \\[-13pt] \nonumber
\end{align}	
where an additional attenuation of 3dB to the reflected path of reflecting element $m$  is assumed, if it is active. This attenuation is considered in the simulations in order to capture the hardware limitations of the utilized RIS prototype~\cite{heinrichs2023RIS}.

\section{Experimental Results}
This section presents and analyzes the results obtained from the experimental measurement campaign described in Section~\ref{sec:exp}.
The objective of the measurements is to experimentally validate the ability of the RIS to shape and control the spatial distribution of the received signal power within the predefined measurement areas using the model-based optimization strategy introduced in the previous section.

For each grid point (i.e., measured spatial position) recorded with the Leica Multistation, the corresponding RIS configuration was computed analytically using the geometry-based channel model and subsequently implemented on the RIS prototype. Throughout this procedure, the receiver position was kept fixed, while the RIS configuration was updated individually for each grid point. The complete measurement campaign was conducted in both areas and, within each area, for two distinct receiver locations: one placed relatively close to the RIS and another at a larger distance. This setup enables a systematic investigation of geometric effects on the achievable spatial focusing performance.

All measurements were conducted at a carrier frequency of $5.375\,\mathrm{GHz}$. The RIS was mounted on a tripod at a height of $3.6\,\mathrm{m}$ above the ground, and the Tx--RIS distance was fixed to $0.587\,\mathrm{m}$. The RIS position and orientation were kept constant across all measurements, as shown in Fig.~\ref{fig:ris_far} and Fig.~\ref{fig:floor}.  For each optimized RIS configuration, the received signal power was measured using the spectrum analyzer. The Rx antenna was positioned at a height of $1.1\,\mathrm{m}$.

The resulting heat maps visualize the measured received power over the spatial grid. To aid visualization, the maps are obtained by interpolating the measured values across the grid. The native spacing (resolution) of the underlying grid points was 10 cm. Since the RIS height and the receiver height are fixed, variations along the grid direction pointing away from the RIS correspond predominantly to changes in the elevation angle, whereas variations along the lateral grid direction correspond predominantly to changes in the azimuth angle.

\subsection{Area 1}
\begin{figure}[htbp]
    \centering
    \subfloat[Rx position~1 (near RIS).]{%
        \includegraphics[width=0.95\linewidth]{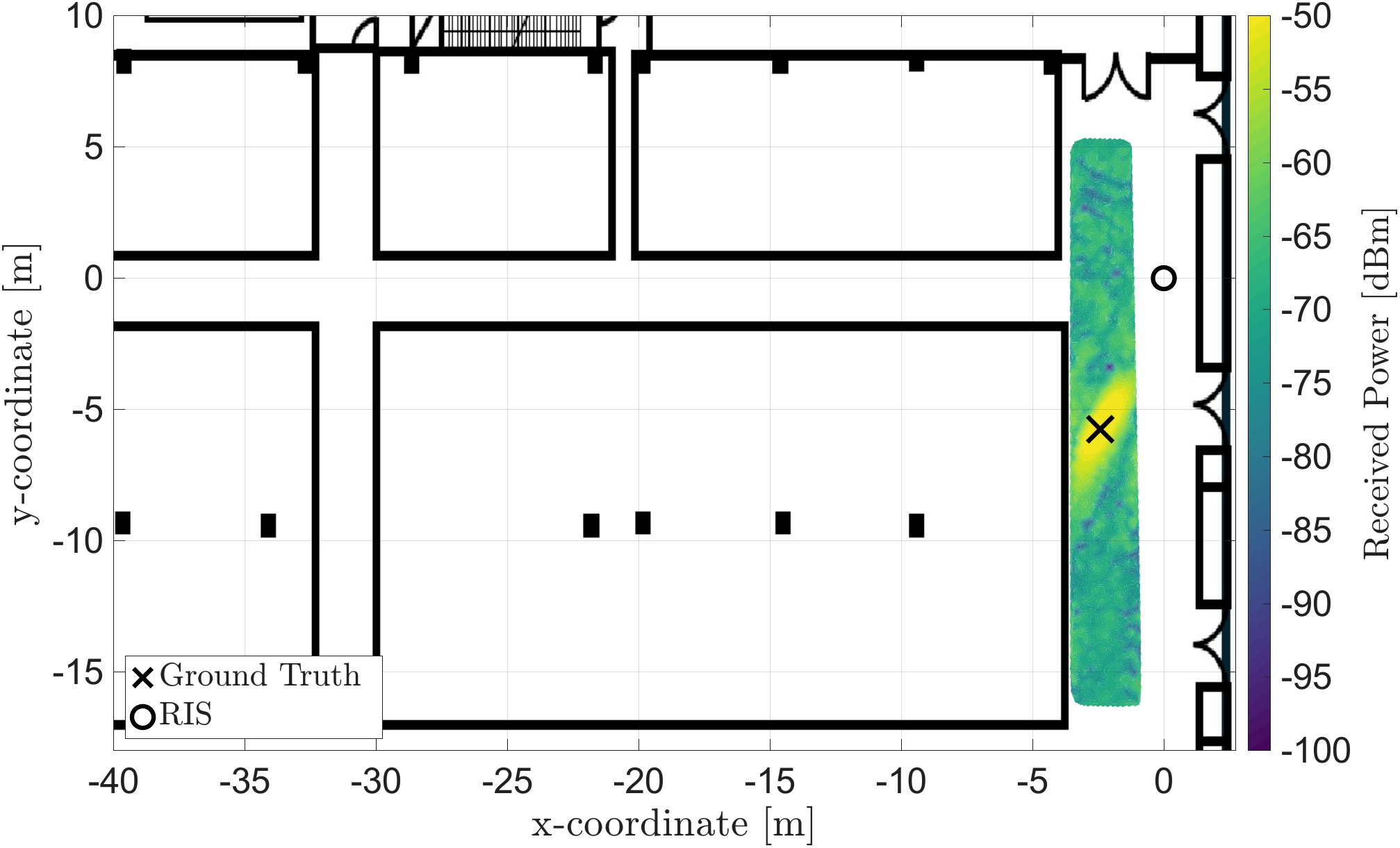}
        \label{fig:area1_ps1}
    }\\
    \subfloat[Rx position~2 (far from RIS).]{%
        \includegraphics[width=0.95\linewidth]{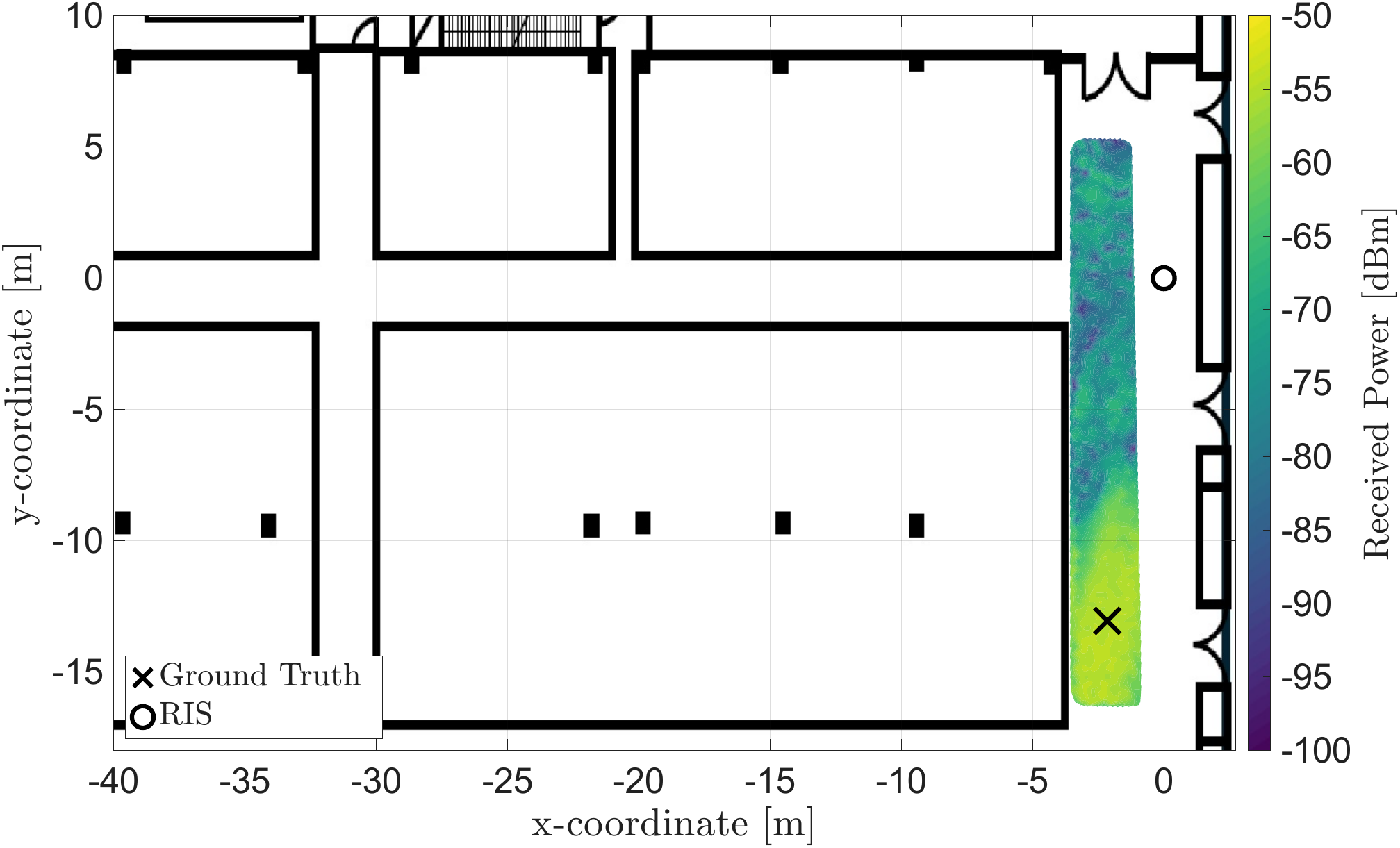}
        \label{fig:area1_ps2}
    }
    \caption{Area~1: measured received power (dBm) heat maps for two receiver positions (near and far), using geometry-based per-grid-point RIS optimization.}
    \label{fig:area1}
\end{figure}

This subsection analyzes the measured spatial distribution of the received signal power in Area~1 for two receiver positions relative to the RIS, as shown in Fig.~\ref{fig:area1}. Fig.~\ref{fig:area1_ps1} depicts the measured received power distribution for receiver position~1, which is located in close proximity to the RIS. The resulting heat map shows a pronounced peak in received power for RIS configurations optimized for grid points near the true receiver position (approximately $-50\,\mathrm{dBm}$). As the spatial offset between the optimized grid point and the receiver increases, the received power drops rapidly by about $20$--$30\,\mathrm{dB}$. This behavior indicates that the analytically derived RIS configurations effectively steer and focus the reflected wavefront toward the intended spatial locations. In addition, the narrow high-power region demonstrates a high degree of spatial selectivity in both grid directions (i.e., primarily azimuth and elevation-angle selectivity), confirming that for short RIS-to-receiver distances a tightly focused reflection pattern can be achieved in a highly reflective industrial environment.

The measured received power distribution for receiver position~2 is shown in Fig.~\ref{fig:area1_ps2}. Here, the receiver is placed farther from the RIS, while the same grid-based optimization procedure is applied. Compared to position~1, the high-power region becomes noticeably broader, and the peak received power decreases to about $-55\,\mathrm{dBm}$, consistent with the increased RIS--receiver distance. Elevated received power is observed not only for RIS configurations optimized for grid points near the receiver location, but also for configurations targeting grid points closer to the RIS. This suggests a larger spread of the reflected energy in the elevation dimension, which can be attributed to propagation geometry and the finite aperture of the RIS. Nevertheless, azimuthal steering remains effective despite the increased RIS-to-receiver distance.

\subsection{Area 2}
This subsection examines the measured spatial distribution of the received power in Area~2 for two receiver locations relative to the RIS. In contrast to Area~1, Area~2 features a more constrained, corridor-like geometry. Therefore, the analysis focuses primarily on steering behavior along the grid direction that mainly corresponds to variations in elevation angle.

\begin{figure}[htbp]
    \centering
    \subfloat[Rx position~1 (near RIS).]{%
        \includegraphics[width=0.95\linewidth]{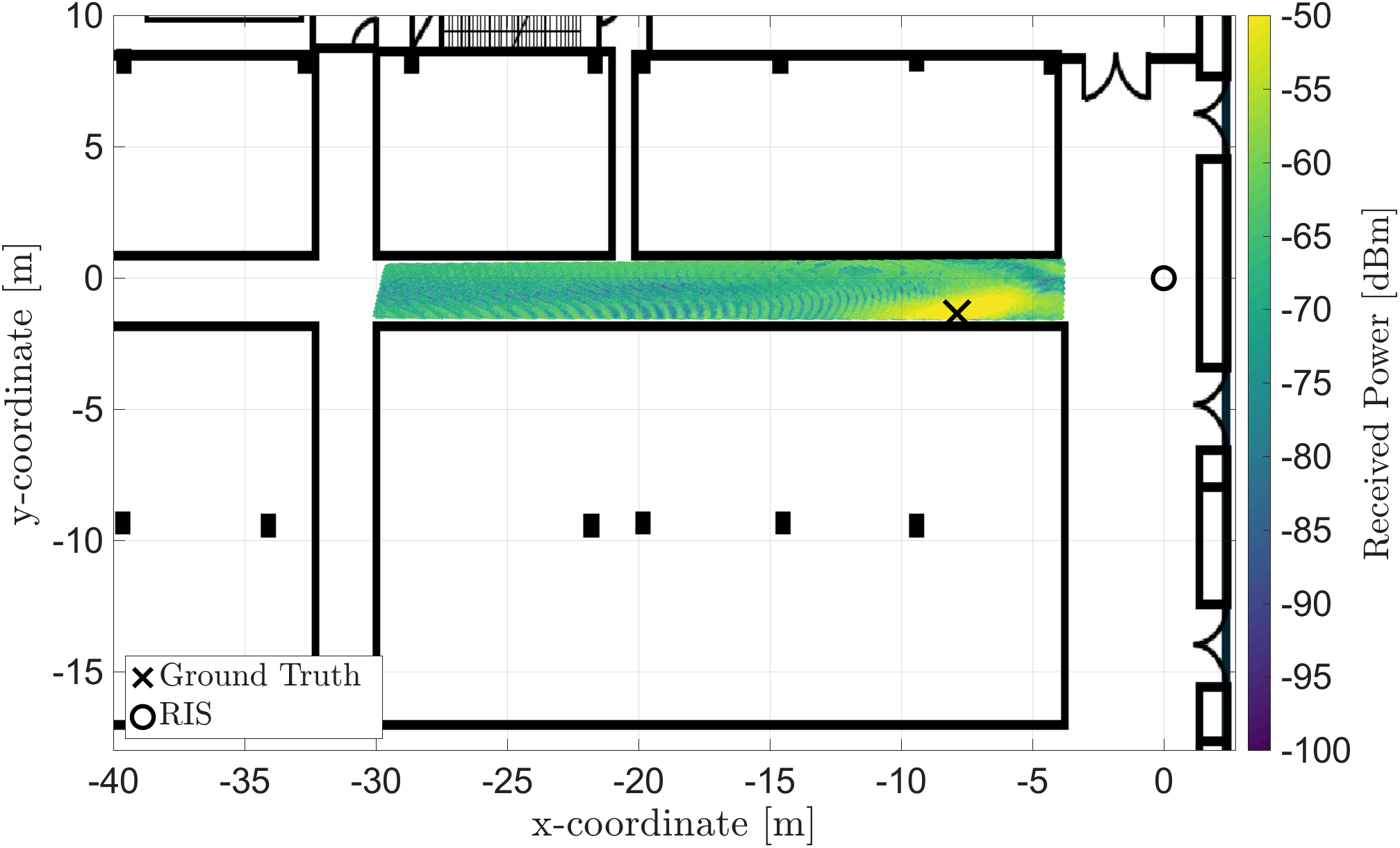}
        \label{fig:area2_ps1}
    }\\
    \subfloat[Rx position~2 (far from RIS).]{%
        \includegraphics[width=0.95\linewidth]{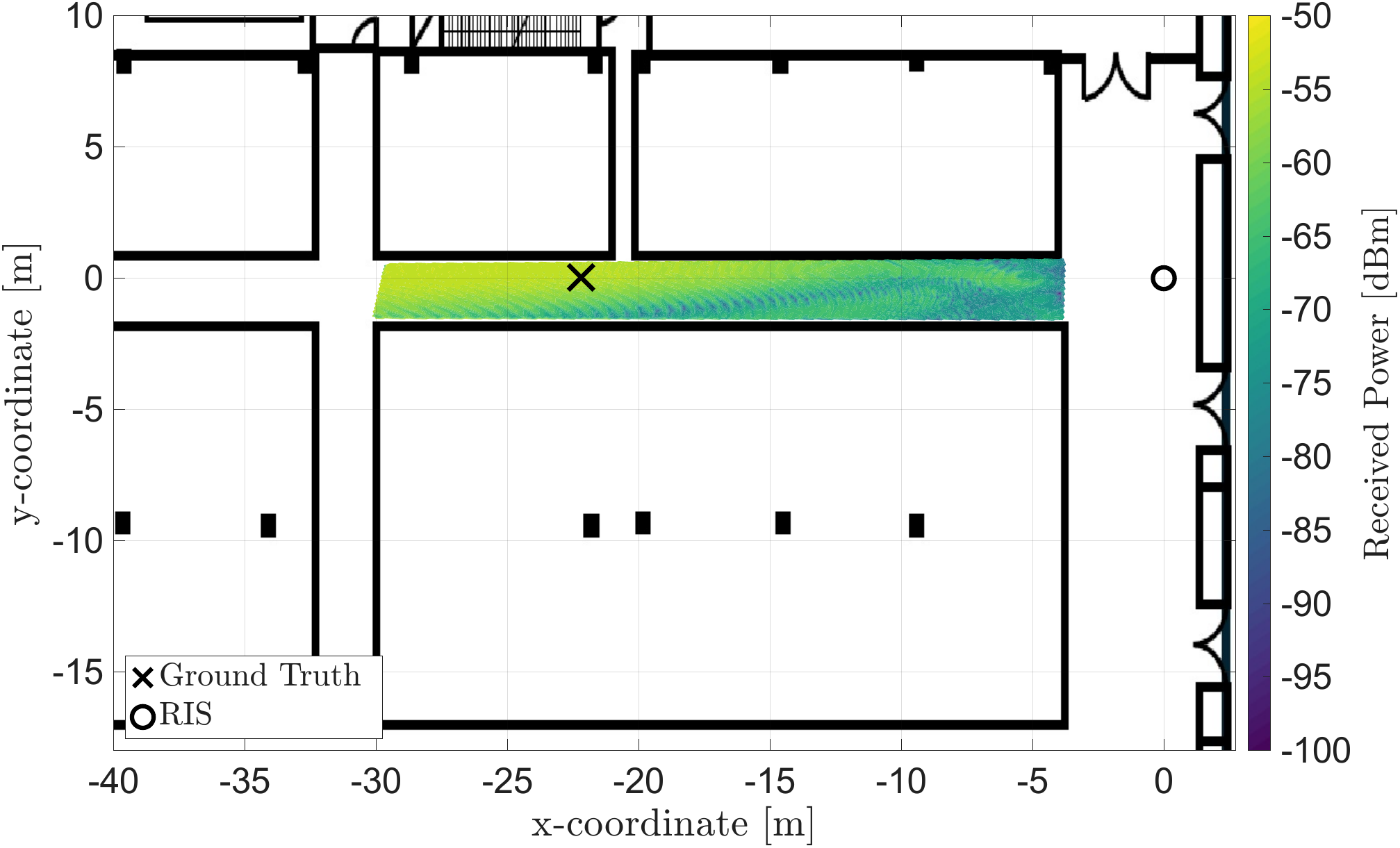}
        \label{fig:area2_ps2}
    }
    \caption{Area~2: measured received power (dBm) heat maps for two receiver positions (near and far), using geometry-based per-grid-point RIS optimization.}
    \label{fig:area2}
\end{figure}

Figure~\ref{fig:area2_ps1} shows the measured received power distribution for receiver position~1, located close to the RIS. The heat map exhibits a clear localization of received power around the grid points corresponding to the true receiver location, with a peak level of approximately $-50\,\mathrm{dBm}$. As the spatial offset between the optimized grid point and the actual receiver position increases, the received power decays rapidly by $20$--$30\,\mathrm{dB}$. This observation is consistent with the behavior in Area~1 and confirms that, at short RIS-to-receiver distances, the analytically computed configurations produce a tightly focused reflection pattern that closely follows the intended grid-point targets.

The received power distribution for receiver position~2 is presented in Fig.~\ref{fig:area2_ps2}. Here, the receiver is located farther from the RIS, leading to a reduced peak received power of about $-55\,\mathrm{dBm}$ and a noticeably broader high-power region, particularly along the grid direction associated with elevation-angle variations. Elevated received power occurs not only for configurations optimized for grid points near the receiver, but also for configurations targeting points closer to the RIS. This indicates increased spreading of the reflected wavefront, which can be attributed to propagation geometry and the limited effective aperture of the RIS at larger distances.

\subsection{Discussion}

The measurements show that RIS configurations computed from a geometry-based model can impose a structured and spatially selective received-power distribution in a large industrial hall. In both measurement areas, configurations optimized near the true receiver position yield a clear power maximum, while targeting offset grid points produces a rapid power reduction by 20--30\,dB. This indicates that geometric steering remains effective despite strong multipath and metallic scattering. The observed patterns deviate from an ideal focus. Peak broadening and reduced contrast are consistent with practical non-idealities of the setup, including binary phase control, finite effective aperture, finite grid resolution, and residual model mismatch. A consistent trend is that increasing the RIS--receiver distance reduces selectivity along the grid direction associated with elevation-angle variations, whereas azimuthal steering remains comparatively robust; this aligns with aperture- and geometry-driven angular compression at larger ranges and the effect of phase discretization.

Taken together, the results suggest that geometry-driven configuration can provide spatial structure that is stable enough to be informative beyond pure power maximization. In particular, the strong contrast between configurations targeting nearby versus displaced grid points implies location-dependent signatures that could be exploited for coarse localization, e.g., by matching measured power responses to a precomputed map or by using power gradients to refine candidate positions. Establishing this link rigorously will require quantifying resolution and repeatability over time and across deployments, and evaluating localization performance under realistic dynamics.

\section{Conclusion}
This paper presented a measurement-based validation of geometry-driven RIS beam steering in a large-scale industrial production hall characterized by strong multipath propagation and metallic scattering. A RIS prototype was illuminated by four patch antennas mounted directly in front of the surface. Using analytically derived phase configurations, spatially selective and repeatable beam steering was demonstrated. This was achieved without relying on measurement feedback or environment-specific model refinement. The measured field distributions closely follow the intended geometric targeting, confirming the practical applicability of model-based RIS optimization under realistic industrial conditions. While increasing the RIS--receiver distance leads to reduced elevation resolution due to fundamental geometric and aperture limitations, robust azimuthal steering performance is preserved. These results highlight the suitability of RIS-assisted systems for industrial localization tasks and support their integration into future industrial wireless networks.

\bibliographystyle{IEEEtran}
\bibliography{bibliography}

@INPROCEEDINGS{weinberger2023geo,
  author={Weinberger, Kevin and Tewes, Simon and Heinrichs, Markus and Kronberger, Rainer and Sezgin, Aydin},
  booktitle={Proc. WSA \& SCC}, 
  title={{RIS}-Based Channel Modeling and Prototypical Validation}, 
  year={2023},
  volume={},
  number={},
  pages={1-6},
  doi={}}

@INPROCEEDINGS{heinrichs2023RIS,
  author={Heinrichs, Markus and Sezgin, Aydin and Kronberger, Rainer},
  booktitle={Proc. ISAP}, 
  title={Open Source Reconfigurable Intelligent Surface for the Frequency Range of 5 {GHz} {WiFi}}, 
  year={2023},
  volume={},
  number={},
  pages={1-2},
  doi={10.1109/ISAP57493.2023.10389095}}

@ARTICLE{renzo2020RIS,
  author={Di Renzo, Marco and Zappone, Alessio and Debbah, Merouane and Alouini, Mohamed-Slim and Yuen, Chau and de Rosny, Julien and Tretyakov, Sergei},
  journal={IEEE J. Sel. Areas Commun.}, 
  title={Smart Radio Environments Empowered by Reconfigurable Intelligent Surfaces: How It Works, State of Research, and The Road Ahead}, 
  year={2020},
  volume={38},
  number={11},
  pages={2450-2525},
  doi={10.1109/JSAC.2020.3007211}}

@ARTICLE{wu2020RIS,
  author={Wu, Qingqing and Zhang, Rui},
  journal={IEEE Commun. Mag.}, 
  title={Towards Smart and Reconfigurable Environment: Intelligent Reflecting Surface Aided Wireless Network}, 
  year={2020},
  volume={58},
  number={1},
  pages={106-112},
  doi={10.1109/MCOM.001.1900107}}

@ARTICLE{weinberger2022RISRAN,
  author={Weinberger, Kevin and Ahmad, Alaa Alameer and Sezgin, Aydin and Zappone, Alessio},
  journal={IEEE Trans. Wirel. Commun.}, 
  title={Synergistic Benefits in {IRS}- and {RS}-Enabled {C-RAN} With Energy-Efficient Clustering}, 
  year={2022},
  volume={21},
  number={10},
  pages={8459-8475},
  doi={10.1109/TWC.2022.3166393}}

@misc{umra2025hardwareefficientcognitiveradarmultitarget,
      title={Hardware-Efficient Cognitive Radar: Multi-Target Detection with {RL}-Driven Transmissive {RIS}}, 
      author={Adam Umra and Aya Mostafa Ahmed and Stefan Roth and Aydin Sezgin},
      year={2025},
      eprint={2509.14160},
      archivePrefix={arXiv},
      primaryClass={eess.SP},
      url={https://arxiv.org/abs/2509.14160}, 
}

@ARTICLE{magbool2025RIS,
  author={Magbool, Ahmed and Kumar, Vaibhav and Wu, Qingqing and Di Renzo, Marco and Flanagan, Mark F.},
  journal={IEEE Open J. Commun. Soc.}, 
  title={A Survey on Integrated Sensing and Communication With Intelligent Metasurfaces: Trends, Challenges, and Opportunities}, 
  year={2025},
  volume={6},
  number={},
  pages={7270-7318},
  doi={10.1109/OJCOMS.2025.3594049}}

@misc{FraunhoferIPT_5GIndustryCampusEurope,
  author       = {{Fraunhofer Institute for Production Technology (IPT)}},
  title        = {{5G}-Industry Campus Europe},
  year         = {n.d.},
  howpublished = {\url{https://www.ipt.fraunhofer.de/de/profil/hallenrundgang.html}},
  note         = {Online; accessed 13 January 2026},
  institution  = {Fraunhofer Institute for Production Technology}
}

@inbook{goldsmith2005WC, place={Cambridge}, title={Wireless Communications}, booktitle={Wireless Communications}, publisher={Cambridge University Press}, author={Goldsmith, Andrea}, year={2005}}

@ARTICLE{luvisotto2019Indust,
  author={Luvisotto, Michele and Pang, Zhibo and Dzung, Dacfey},
  journal={Proc. IEEE}, 
  title={High-Performance Wireless Networks for Industrial Control Applications: New Targets and Feasibility}, 
  year={2019},
  volume={107},
  number={6},
  pages={1074-1093},
  doi={10.1109/JPROC.2019.2898993}}

@INPROCEEDINGS{weinberger2024UWBris,
  author={Weinberger, Kevin and Tewes, Simon and Sezgin, Aydin},
  booktitle={Proc. 19th ISWCS}, 
  title={Show Me the Way: Real-Time Tracking of Wireless Mobile Users with {UWB}-Enabled {RIS}}, 
  year={2024},
  volume={},
  number={},
  pages={1-6},
  doi={10.1109/ISWCS61526.2024.10639136}}

@INPROCEEDINGS{weinberger2024ValRIS,
  author={Weinberger, Kevin and Tewes, Simon and Sezgin, Aydin},
  booktitle={Proc. 18th EuCAP}, 
  title={Validating Properties of RIS Channel Models with Prototypical Measurements}, 
  year={2024},
  volume={},
  number={},
  pages={1-5},
  doi={10.23919/EuCAP60739.2024.10501125}}

@ARTICLE{basar2019RISaccess,
  author={Basar, Ertugrul and Di Renzo, Marco and De Rosny, Julien and Debbah, Merouane and Alouini, Mohamed-Slim and Zhang, Rui},
  journal={IEEE Access},
  title={Wireless Communications Through Reconfigurable Intelligent Surfaces},
  year={2019},
  volume={7},
  number={},
  pages={116753-116773},
  doi={10.1109/ACCESS.2019.2935192}}

@Article{noor2023Smart,
AUTHOR = {Noor-A-Rahim, Md. and John, Jobish and Firyaguna, Fadhil and Sherazi, Hafiz Husnain Raza and Kushch, Sergii and Vijayan, Aswathi and O’Connell, Eoin and Pesch, Dirk and O’Flynn, Brendan and O’Brien, William and Hayes, Martin and Armstrong, Eddie},
TITLE = {Wireless Communications for Smart Manufacturing and Industrial {IoT}: Existing Technologies, {5G} and Beyond},
JOURNAL = {Sensors},
VOLUME = {23},
YEAR = {2023},
NUMBER = {1},
ARTICLE-NUMBER = {73},
PubMedID = {36616671},
ISSN = {1424-8220},
DOI = {10.3390/s23010073}
}

@article{mohr2024Matrix,
title = {Design of Matrix Production Systems: New Demands on Factory Planning Methods},
journal = {Procedia Computer Science, Part of special issue: 5th ISM 2023},
volume = {232},
pages = {1972-1981},
year = {2024},
issn = {1877-0509},
doi = {https://doi.org/10.1016/j.procs.2024.02.019},
author = {Joshua Mohr and Niels Schmidtke and Fabian Behrendt}
}

\end{document}